\documentclass[prd,superscriptaddress,amsfonts,amssymb,amsmath,showpacs]{revtex4-2}
\usepackage{bm}
\usepackage{amsfonts}
\usepackage{latexsym}
\usepackage[latin1]{inputenc}
\usepackage{graphicx}
\usepackage{amsmath}
\usepackage{palatino}
\usepackage{mathpazo}
\usepackage{natbib}
\usepackage{textcomp}
\usepackage{float}
\usepackage{booktabs}
\usepackage{dcolumn}
\usepackage{booktabs}
\usepackage{multirow}
\usepackage{hyperref}
\hypersetup{colorlinks=true,citecolor=red,linkcolor=blue,urlcolor=blue}
\usepackage{amsmath}
\usepackage{xcolor}
\usepackage{orcidlink}
\usepackage[caption=false]{subfig}
\usepackage{commath}

\def\jnl@style{\it}
\def\aaref@jnl#1{{\jnl@style#1}}

\def\aaref@jnl#1{{\jnl@style#1}}

\def\aj{\aaref@jnl{AJ}}                   
\def\apj{\aaref@jnl{ApJ}}                 
\def\apjl{\aaref@jnl{ApJ}}                
\def\apjs{\aaref@jnl{ApJS}}               
\def\apss{\aaref@jnl{Ap\&SS}}             
\def\aap{\aaref@jnl{A\&A}}                
\def\aapr{\aaref@jnl{A\&A~Rev.}}          
\def\aaps{\aaref@jnl{A\&AS}}              
\def\mnras{\aaref@jnl{Mon.~Not.~Roy.~Astron.~Soc.}}             
\def\prd{\aaref@jnl{Phys.~Rev.~D}}        
\def\prc{\aaref@jnl{Phys.~Rev.~C}}  
\def\prl{\aaref@jnl{Phys.~Rev.~Lett.}}    
\def\qjras{\aaref@jnl{QJRAS}}             
\def\skytel{\aaref@jnl{S\&T}}             
\def\ssr{\aaref@jnl{Space~Sci.~Rev.}}     
\def\zap{\aaref@jnl{ZAp}}                 
\def\nat{\aaref@jnl{Nature}}              
\def\aplett{\aaref@jnl{Astrophys.~Lett.}} 
\def\apspr{\aaref@jnl{Astrophys.~Space~Phys.~Res.}} 
\def\physrep{\aaref@jnl{Phys.~Rep.}}      
\def\physscr{\aaref@jnl{Phys.~Scr}}       
\def\commat{\aaref@jnl{Comm.~Math.~Phys.}}              
\def\science{\aaref@jnl{Science}}               
\def\cqg{\aaref@jnl{Classical Quant.~Grav.}}            
\def\jpcs{\aaref@jnl{JPCS}}                                     
\def\ijmpd{\aaref@jnl{Int.~J.~Mod.~Phys.~D}}                    
\def\grg{\aaref@jnl{Gen.~Relat.~Gravit.}}               
\def\rpp{\aaref@jnl{Rep.~Prog.~Phys.}}          
\def\npa{\aaref@jnl{Nucl.~Phys.~A}}        
\def\lrr{\aaref@jnl{Living Rev.~Rel.}}                   
\def\jcap{\aaref@jnl{J.~Cosmology Astropart.~Phys.}}    
\def\rmp{\aaref@jnl{Rev.~Mod.~Phys.}}   
\def\epjc{\aaref@jnl{Eur.~Phys.~J.~C}}

\allowdisplaybreaks[1]

\begin{document}

\color{black}       

\title{Black Holes and Wormholes Beyond Classical General Relativity} 

\author{A. S. Agrawal \orcidlink{0000-0003-4976-8769}}
\email{asagrawal.sbas@jspmuni.ac.in}
\affiliation{Department of Mathematics, Jayawant Shikshan Prasarak Mandal University, Pune, India.}
\author{Sergio Zerbini \orcidlink{0000-0002-0991-1084}}
\email{sergio.zerbini@unitn.it}
\affiliation{Dipartimento di Fisica, Universit\`a di Trento,Via Sommarive 14, 38123 Povo (TN), Italy}
\author{B. Mishra \orcidlink{0000-0001-5527-3565}}
\email{bivu@hyderabad.bits-pilani.ac.in}
\affiliation{Department of Mathematics, Birla Institute of Technology and Science-Pilani,\\ Hyderabad Campus, Hyderabad-500078, India.}

\begin{abstract}
\textbf{Abstract}:
In the paper only Static Spherically Symmetric space-times in four dimension are considered within  modified gravity models.
The  non-singular static metrics, including black holes not admitting a de Sitter core in the centre and traversable wormholes, are reconsidered within a class of higher-order $F(R)$, satisfying the constraints $F(0)=\frac{dF}{dR}(0)=0$.   Furthermore, making use of the so called effective field theory formulation of gravity, the quantum corrections to Einstein-Hilbert  action due to higher-derivative terms related to curvature invariants are investigated.  
  In particular, in the case of  Einstein-Hilbert action plus cubic curvature Goroff-Sagnotti contribution,   the second order correction in the Goroff-Sagnotti coupling constant  is computed. 
In general, it is shown that the effective metrics, namely Schwarschild expression plus small quantum corrections are related to  black holes, and  not to traversable wormholes. In this framework, within the approximation considered, the resolution of singularity for $r=0$ is not accomplished.  The related properties of these solutions are investigated.
\end{abstract}

\maketitle
\section{Introduction}
In Einstein's General Relativity (GR), the action, an integral summarizing the dynamics of spacetime and matter, is proportional to the Ricci scalar $R$. $f(R)$ gravity generalizes this by making the action as a function of $R$. Pure $R^2$ gravity is considered to be one of the simplest candidates for modified gravity, while the traditional Einstein-Hilbert term is suppressed. The theory was first considered in the 1960s by Buchdahl as a parsimonious prototype of higher-order gravity that possesses an additional symmetry-scale invariance \cite{Buchdahl1962}. There has been a recent surge of interest in the pure $R^2$ action within a larger context of modified gravity \cite{Stelle1977, Stelle1978, Defe, faraoni, Capozziello2011, Nojiri2011,  Clifton2012, Alvarez-Gaume2016, Alvarez2018}. Pure $R^2$ gravity is unique in the sense that it is both ghost-free and scale-invariant \cite{Kounnas2015}.

The quest for a consistent and predictive theory of quantum gravity remains one of the key challenges in modern physics \cite{Loll2022, Rovelli2010}. Overcoming these challenges would represent significant progress towards a viable theory of quantum gravity. This will be capable of addressing unresolved issues such as the information paradox and the singularity problem that arises in classical and semi-classical approaches to gravitational phenomena. One intriguing proposal within the framework of superstring theory \cite{Polchinski2011} is the fuzzball proposal \cite{Lunin2002a, Lunin2002b}, which offers a potential resolution to these issues. According to this proposal, a black hole (BH) is envisioned as a massive object composed of a vast number of microscopic strings, each with a minimal length extension on the order of the Planck scale. However, it is important to mention that the fuzzball proposal is not the only self-consistent and robust alternative for the generalization of BH physics incorporating quantum gravitational effects. Another important class of BHs can be derived within the framework of higher-derivative theories and non-local gravity \cite{Calmet2022}. These models address issues arising from the combination of quantum field theory and gravitation such as non-renormalizability, non-unitarity by incorporating higher-derivative terms in the Lagrangian of gravitational interaction. Interestingly, these additional terms can eliminate the undesirable features that impact the standard quantization of Einstein's GR and may lead to potentially detectable effects in various physical phenomena.

BHs are incredibly interesting for a variety of reasons. Stephen Hawking's groundbreaking insight that BHs are not actually black but have a radiation spectrum similar to that of a black body makes them an ideal environment for studying the relationship between quantum mechanics, gravity, and thermodynamics. This has led to the concept of BH entropy, which has garnered significant attention over the past years. The discovery of gravitational waves from binary BH systems and the multimessenger signals from the first observation of a collision between two relativistic neutron stars have significantly expanded our understanding of the most massive and compact objects of the Universe. Unlike other known interactions in nature, these events have demonstrated the fundamental role of high gravity effects. GR predictions have been largely confirmed, and data from LIGO (Laser Interferometer Gravitational-Wave Observatory) \cite{LIGO1, LIGO2, LIGO3, LIGO4} has confirmed the existence of gravitational waves. Gravitational wave signals have been associated with the largest compact objects in nature, namely Kerr BHs, within the margins of experimental error. However, the potential existence of less extreme compact objects cannot be discounted. It is worth noting that a Kerr BH is a stationary vacuum solution of GR characterized by a singularity. This has prompted extensive research into finding a feasible alternative.

We note that BHs produce a ringdown waveform that is entirely determined by Quasinormal modes (QNMs), which are dependent solely on the mass and angular momentum of the BH. Therefore, the potential existence of additional ``echoes'' in the ringdown waveform could indicate the absence of an event horizon. This implies that horizonless compact objects (HCOs) such as gravastars ~\cite{grav, grav2}, boson stars ~\cite{bs}, or other exotic compact objects (ECOs) ~\cite{Mark, ECO1, ECO2, Conk, Ma, ECO3, ECO4} cannot be ruled out as an alternative to BHs. Lambiase et al. \cite{Lambiase2024} examined how the quantum-gravity correction parameter $c_6$ at the third-order curvature affects black hole shadows and weak deflection angles. They achieved this by studying the photon sphere and using the finite-distance version of the Gauss-Bonnet theorem (GBT).

The paper is structured as follows: In Sec. \ref{NSBH}, we introduce non-singular BHs and wormholes (WH) in $F(R)$ modified gravity. Sec. \ref{WHR2G} is dedicated to the WH solution in $R^2$ gravity, specifically a new spherically symmetric WH solution for $R^2$. In Sec. \ref{QGCCSS},  making use of the effective field theory formulation of gravity, we will revisit the quantum gravitational corrections to the classical vacuum solution of GR. In Sec. \ref{TLCBH}, we discuss the two-loop quantum-corrected BH related to the Goroff-Sagnotti contribution. Sec. \ref{GR_MGM} covers $F(R)$ modified gravitational models. In Sec. \ref{PRS}, we will explore the idea of a photon ring and shadow linked to a generic Static Spherically Symmetric (SSS) compact object, such as a BH or a WH. Finally, in Sec. \ref{Conclusion}, we provide commentary on our results and offer perspectives on future research. In the paper we use the conventions $c=\frac{h}{2\pi}=1$, as well as the Newton constant $G_N=1$.

\section{Non singular black holes and Wormholes in \texorpdfstring{$F(R)$}{} modified gravity}\label{NSBH} 

In this section, we discuss the nonsingular BH and WH solutions within a particular class of modified $F(R)$-gravity theories\cite{Dup, calza}. We write down the equations of motion for $F(R)$- gravity \cite{faraoni, odi, Defe, capo} in a vacuum, namely with vanishing stress tensor matter as, 
\begin{equation}
F_R(R) G_{\mu \nu}=\frac{1}{2}\left(F(R)- RF_R(R) \right)g_{\mu \nu}+\left(\nabla_\mu  \nabla_\nu-g_{\mu \nu} \nabla^2 \right)F_R(R)\,.
\label{fr}
\end{equation}
Here, $G_{\mu \nu}$ is the usual Einstein tensor, and $F_R=\frac{d F}{dR} $. We consider the class of modified gravity models such that (on the shell) $F(R^*)=0$ and $F_R(R^*)=0$, where $R^*$ is a constant Ricci scalar.
As a result, taking into account \eqref{fr}, any spherically symmetric static metric (the ones we are interested in) such that identically leads to a constant Ricci scalar is a solution of the above equations of motion within our special class of models. For the sake of simplicity, we will assume $R^*=0$.\\

The Ricci tensor, $R_{\mu \nu}$ takes the form
\begin{equation}
    R_{\mu \nu}=\frac{\partial \Gamma ^{\gamma}_{\mu \nu}}{\partial x^{\gamma}}-\frac{\partial \Gamma^{\gamma}_{\mu \gamma}}{\partial x^{\nu}}+\Gamma^{\lambda}_{\mu \nu}\Gamma^{\gamma}_{\lambda \gamma}-\Gamma^{\lambda}_{\gamma\mu}\Gamma^{\gamma}_{\nu \lambda}.  
\end{equation}

The Christoffel symbol in the above equation is
\begin{equation}
\Gamma^{\gamma}_{\mu \nu}=\frac{1}{2}g^{\gamma\sigma}\left(\frac{\partial g_{\sigma\nu}}{\partial x^{\mu}}+\frac{\partial g_{\sigma\mu}}{\partial x^{\nu}}-\frac{\partial g_{\mu \nu}}{\partial x^{\sigma}}\right),    
\end{equation}

where the $R=g^{\mu \nu}R_{\mu \nu}$ is the Ricci scalar.

The Kretschmann scalar, $K=R^{\mu \nu \gamma \sigma}R_{\mu \nu \gamma \sigma}$ measures the curvature of spacetime near a black hole and is often used to analyze the nature of singularities. For a more detailed and rigorous understanding of this topic, one can refer to academic papers and research articles that delve into the mathematical and physical implications of the Kretschmann scalar in the context of singularities \cite{Kretschmann}. These papers offer in-depth discussions and clarity on the behavior of spacetime curvature at the singularity of a black hole.

The most simple and important example is the scale invariant quadratic gravity $F(R)=R^2$, investigated in Ref. \cite{Max}, and its effective one-loop correction, investigated in Ref. \cite{max2}, 
\begin{equation}
F(R)=R^2-bR^2\ln\left( \frac{R^2}{\mu^2} \right) \,.
\label{fr1}
\end{equation}
Other examples are possible modifications of GR, namely
\begin{equation}
F(R)=M^2R-M^2R_0\ln\left(1+\frac{R}{R_0} \right) \,.
\label{fr11}
\end{equation}
For very large $R_0$, this model reduces to quadratic gravity plus a higher-order term in the Ricci scalar. For small values of $R_0$, this is GR plus small corrections. In these examples, the Lagrangian and its first derivative with respect to the Ricci scalar go to zero when $R=0$. For other examples, see Ref. \cite{calza, Giacomozzi}.

\subsection{A non-singular Black Hole}
Here, we recall the BH solution found in Ref. \cite{bertipagani}. Let us consider the quite general SSS metric of the kind 
\begin{equation}
ds^2=-A(r)dt^2+\frac{ r dr^2}{(r-\ell)A(r)}+ r^2 dS^2\,,
\label{r1a}
\end{equation}
where $A(r)$ is a function of the radial coordinate $r$ only and $\ell$ is a positive length parameter, eventually on the Planckian size. The associated Ricci scalar reads
\begin{equation}
R=\frac{1}{2 r^2}\left(2(\ell r-r^2)A''+ (7\ell-8r)A'-4(A-1) \right)\,,
\label{r2a}
\end{equation}

where the prime denotes a derivative with respect to the radial coordinate $r$. Imposing $R=0$, one gets the second-order linear differential equation as,  
\begin{equation}
2(\ell r-r^2)A''+ (7\ell-8r)A'-4(A-1)=0\,.
\label{r3}
\end{equation}

On solving, we obtain
\begin{equation}
 A(r)=1-\frac{C}{r}-\frac{3C \ell}{r^{5/2}}\left( \sqrt{r-\ell}\ln\left(\sqrt{r-\ell}+\sqrt{r}\right)-\sqrt{r} \right)+
   \frac{Q\sqrt{r-\ell}}{r^{5/2}} \,,
\label{r33}
\end{equation}
where $C$ and $Q$ are two constants of integration. In the following, we make the choice $C>0$, keeping $Q$ real. The horizons of the metric (\ref{r1a}) are the non-negative real zeros of $A(r)\frac{r-\ell}{r}$. There exists the trivial zero $r=\ell$ and the possible zeroes related to $A(r)$.

The regularity of the metric  \eqref{r1a}  can be understood following the remarks contained in \cite{Sebastiani}. The argument proceeds as follows. Introduce the new radial coordinate $\sigma$ defined by
\begin{equation}
 r^2=\sigma^2+   \ell^2\,.
\end{equation}
Thus, one has
\begin{equation}
ds^2=-A(r(\sigma))dt^2+ \frac{d\sigma^2}{A(r(\sigma))}
    \left(  1+\frac{\ell}{\sqrt{\sigma^2+\ell^2}} \right)
+(\sigma^2+\ell^2)dS^2\,.
\end{equation}

In this form, the regularity is manifest since $r=0$ is excluded as soon as $\ell>0$.

When $\ell$ is not vanishing but sufficiently small, the situation is quite different since $A(\ell)=1+\frac{2C}{\ell}>0$. Thus, for $C>0$, a Cauchy horizon is always present. However, a suitable choice of the parameter may eliminate the Cauchy horizon
\cite{bertipagani}.
\section{Wormhole solution in \texorpdfstring{$R^2$}{} gravity}\label{WHR2G}

First, we recall the formalism we are going to use,  see \cite{noi} for details.  A generic spherically symmetric space-time in isotropic coordinates reads
\begin{equation}
ds^2=- F(\rho)dt^2+G(\rho)\left(d\rho^2+\rho^2 dS^2 \right)\,.   
\label{a1}
\end{equation}

The areal radius is $r=\rho\sqrt{G(\rho)}$. The invariant which defines the presence of the apparent horizon is 
\begin{equation}
 \chi=\frac{\mathring{r}^2}{G}= \frac{(2G+\rho \mathring{G})^{2}}{4G^{2}}. 
\end{equation}

The over-ring represents the derivative with respect to $\rho$. Thus, the horizon is located where $\chi$ is vanishing, namely
\begin{equation}
 2G_H+\rho_H\mathring{G}_H=0\,.   
\label{a2}
\end{equation}

In order to check if one is dealing with a BH or WH, one has to compute the Kodama energy associated with a classical test particle with classical action $I$,
\begin{equation}
 \omega_K=-K^\mu \partial_t I\,.    
\end{equation}

The Kodama vector \cite{noi}, $K^\mu=\left(\dfrac{\mathring{r}}{\sqrt{FG}},0,0,0   \right)$. As a result, Kodama energy is
\begin{equation}
 \omega_K=\frac{E}{2G\sqrt{F}}(2G+\rho \mathring{G})\,.   
\label{a3}
\end{equation}

On the horizon, the numerator is vanishing. In the case of a BH, the Kodama energy is non-vanishing, and $F(\rho)$ must have a double zero in $\rho$. In the case of WH, the Kodama energy is vanishing, and $F_H$ is different from zero. As a check, we recall that the Schwarzschild solution in isotropic coordinates corresponds to 
\begin{equation}
 F(\rho)=\left( \frac{4\rho-r_s}{4\rho+r_s} \right)^2\,, \quad G(\rho)=\left(\frac{1}{4\rho}\right)^4(4\rho+r_s)^4\,.   
\label{sc}
\end{equation}

The horizon is determined by Eq. (\ref{a2}), namely $\rho_H=\frac{r_s}{4}$. One is dealing with a BH, because $F(\rho)$ has a
double zero in $\rho_H$, and the Kodama energy is non-vanishing, namely $\omega_K=E$.

\subsection{A New Spherically Symmetric Wormhole Solution for \texorpdfstring{$R^2$}{}}

We consider the following form for SSS space-time:
\begin{equation}
ds^2=-A(r)dt^2+\frac{  dr^2}{B(r)}+ r^2 dS^2\,,
\label{r1b}
\end{equation}
where both $A(r)$ and $B(r)$ are functions of the radial coordinate only.
The Ricci scalar is derived as
\begin{equation}
R=R_1+R_2\,,\quad R_1=\frac{2}{r}(1-B-rB')\,,\quad R_2=\frac{B}{A}\left(-A''+\frac{A'^2}{2A}-\frac{A'B'}{2B} -\frac{2A'}{r}  \right).    
\end{equation}

Furthermore, we assume a Schwarzschild-like form for $B$, namely
\begin{equation}
 B=1-\frac{r_s}{r}\,, \quad B'=\frac{r_s}{r^2}\,, \quad B''=-\frac{2r_s}{r^3}\,,   
\end{equation}
where $r_s$ is a constant. As a result, $R_1=0$. Next, we take 
\begin{equation}
 A=a_1(\lambda+\sqrt{B})^2\,,   
\end{equation}
with $a_1$ and $\lambda$ dimensionless constants. Since $B=1-\frac{r_s}{r}$, a direct calculation gives $R_2=0$, for every value of $a_1$ and $b_1$. As a consequence, taking
\begin{equation}
 a_1=\frac{1}{(1+\lambda)^2}\,,   
\end{equation}
one has the following: asymptotically flat metric 
\begin{equation}
ds^2=-\left(\frac{\lambda+\sqrt{1-\frac{r_s}{r}}}{1+\lambda}\right)^2   dt^2+\frac{  dr^2}{1-\frac{r_s}{r}}+ r^2 dS^2\,,
\label{r2b}   
\end{equation}
and this metric is a solution of modified gravity models with $F(0)=0$, and $F_R(0)=0$. 
Let us study the properties of this solution. The above metric has $B=1-\frac{r_s}{r}$, and $A= 1-\frac{r_s}{(1+\lambda)^2 r}+\frac{2\lambda}{(1+\lambda)^2}\left(\sqrt{1-\frac{r_s}{r}}-1 \right)$. 

We recall that in the gauge (\ref{r1b}), the Kodama energy reads $\omega_K=\sqrt{\frac{B}{A}}E$. At the horizon (if it exists), $B_H=0$, with $B'_H$ not vanishing. In the BH case, also $A_H=0$, $A'_H \neq 0$. In the WH, case $A_H>0$ \cite{noi}.
The horizon reads $r_H=r_s$. Thus, $A(r_s)=\frac{\lambda^2}{(1+\lambda)^2}>0$, and one has to deal with a traversable WH. When $\lambda=0$, one recovers
the Schwarzschild BH. 

For large $r$, one has $A=1-\frac{r_s}{(1+\lambda)r}+\mathcal{O}\left(\frac{r_s^2}{r^2}\right)$. As a consequence, we may put $2M=\frac{r_s}{1+\lambda}$. If $\lambda$ is very small, one has a WH which mimics the Schwarzschild BH.  It should be noted that the quasi-local Misner -Sharp mass is $M_{MS}=\frac{r_s}{2}$. Furthermore, it is easy to show that in isotropic coordinates, the solution (\ref{r2b}) reads  
\begin{equation}
 ds^2=-\frac{1}{(1+b_1)^2}\left(b_1+\frac{4\rho-r_s}{4\rho+r_s}  \right)^2 dt^2+\left( 1+\frac{r_s}{4\rho} \right)^4\left(d\rho^2+\rho^2dS^2  \right) \,.
\end{equation}

Thus,  it does not belong to the Buchdahl class discussed in Appendix \ref{BWS}. A similar WH solution has been obtained in another gravitational model; Ref. \cite{DadVisser}.  

\section{Quantum gravitational corrections to classical Schwarzschild solution}\label{QGCCSS}
The quantum gravitational adjustments to the emission of Hawking particles concept was first introduced by Calmet et al. \cite{calmet} using a Schwarzschild metric, akin to Hawking's original work \cite{Hawking1976}. Expanding on this, they incorporated established quantum gravitational corrections to the metric and discovered that the quantum amplitude for the emission of a Hawking particle is influenced by quantum hair, which refers to the quantum gravitational correction to the classical Schwarzschild background. In this Section, we shall revisit the quantum gravitational corrections to the classical vacuum solution of GR.

\subsection{The one-loop corrections}
Now, let us consider the one-loop gravitational corrections reported in  \cite{Petru} and references therein. One has a metric in isotropic coordinates  (\ref{a1}) and in suitable units
\begin{equation}
 F(\rho)=\left(1-\frac{m}{2\rho}+ \frac{31 m}{30\pi \rho^3}\right)^2 \left(1+\frac{m}{2\rho} -\frac{31 m}{30\pi \rho^3}\right)^{-2}\,, 
 \quad G(\rho)=\left(1+\frac{m}{2\rho}-  \frac{7 m}{30\pi \rho^3}\right)^4\,.
\end{equation}

Now, the horizon is shifted a little by the quantum correction and can be determined again by making use of the equation (\ref{a2}). As a result, one has
\begin{equation}
 1-\frac{m}{2\rho_H}+ \frac{35 m}{30\pi \rho_H^3}=0\,.  
\end{equation}

Since $F(\rho_H)>0$, namely, is not vanishing at the horizon, one is dealing with a traversable WH, and the new horizon becomes the WH throat. We recall that for static WH, the Hawking radiation is absent \cite{noi}.  Furthermore, the presence of the throat implies that singularity in $r=0$ may be resolved. Similar results have been recently obtained in \cite{Arrechea, Beltran} within the so-called improved semi-classical approximation.


\subsection{Effective Field Theory Extension of GR}\label{n2}
Here, we consider the approach of \cite{Cardoso}, where an effective Lagrangian beyond GR has been investigated. We limit to the first correction, parametrized by dimensionless coupling constant $\varepsilon$. The effective Lagrangian is the Einstein-Hilbert plus a quadratic term in Curvature tensor. The complete equations of motion are reported in \cite{Cardoso}. In Ref. \cite{Claudia} a more general effective Lagrangian up to third order in the curvature has been investigated. We will come back to it in the next section. Within a  metric given again by
\begin{equation}
ds^2=-f(r)dt^2+\frac{dr^2}{g(r)}+ r^2 dS^2\,,
\label{r13}   
\end{equation}

in the zero-order $\varepsilon=0$, one is dealing with GR, and the solution is unique and coincides with the Schwarzschild solution. The solutions of the equations of motions up to  the second order in $\varepsilon$ read \cite{Cardoso}
\begin{subequations}
    \begin{equation}
 f(r)=1-\frac{2M}{r}+\varepsilon \left(\frac{2M}{r}\right)^9\left(\frac{11 M}{4r}-2 \right)+\varepsilon^2f_2(r)\,,   
\end{equation}
where
\begin{equation}
 f_2(r)=-\frac{1}{544}\left(\frac{2M}{r}\right)^{19}\left(  198305-231709 \frac{r}{M}+64584 \frac{r^2}{M^2}  \right) \,, \label{f22}
\end{equation}
\begin{equation}
 g(r)=1-\frac{2M}{r}+\varepsilon \left(\frac{2M}{r}\right)^9\left(\frac{67 M}{4r}-9 \right)+\varepsilon^2g_2(r)\,,   
\end{equation}
where
\begin{equation}
 g_2(r)=-\frac{1}{544}\left(\frac{2M}{r}\right)^{19}\left( 2513399-2370222 \frac{r}{M}+554472 \frac{r^2}{M^2}  \right)\,. \label{g22}
\end{equation}
\end{subequations}
Where $M$ be the mass of the black hole.

The horizon may be found solving $g(r_{H})=0$, namely
\begin{equation}
 r_H=2M-  \varepsilon \left(\frac{2M}{r_H}\right)^9\left(\frac{67 M}{4}-9r_H \right)-\varepsilon^2 r_H g_{2}(r_H)\,.
\end{equation}

Looking for a solution perturbatively in $\varepsilon$,
\begin{equation}
 r_H=r_0+\varepsilon r_1+\varepsilon^2 r_2\,,
\end{equation}
one has
\begin{equation}
 r_0=2M\,, \quad r_1=\frac{5}{8}r_0\,,\quad r_2=r_0\left(\frac{135}{64}-g_2(r_0)\right)\,,   
\end{equation}
$r_1$ is correct in agreement with \cite{Cardoso},  the computation of $r_2$ has been done starting from

\begin{equation}
 r_0+  \varepsilon r_1+\varepsilon^2 r_2 =2M-  \varepsilon \left(\frac{2M}{r_H}\right)^9\left(\frac{67 M}{4}-9r_H \right)-\varepsilon^2 r_H g_{2}(r_H)\,.
\end{equation}

In order to check if one is dealing with a BH, then need to evaluate $f(r_H)$. So,
\begin{equation}
 f(r_H)-g(r_H)=\varepsilon^2\left(\frac{35}{8}+f_2(r_0)-g_2(r_0)  \right)\,.   
\end{equation}

Making use of equations (\ref{f22}) and (\ref{g22}), one can obtain
\begin{equation}
f_2(r_0)-g_2(r_0)=-\frac{2380}{544}=-\frac{35}{8}\,.    
\end{equation}

As a consequence, $f(r_H)=0$, and one is dealing again with a BH.

\section{Two-loop quantum corrected  black hole}\label{TLCBH}

Since the one-loop corrections are vanishing in a vacuum, other and different quantum gravitational corrections to the Schwarzschild solution have been proposed \cite{calmet, Alvarez}. In fact, as shown by Goroff and Sagnotti (GS) \cite{GS}, the two-loop quantum correction is not vanishing in a vacuum. As a consequence, one may consider the following effective action consisting in the Einstein-Hilbert term plus the GS higher derivative counter term, an invariant which is cubic in the curvature tensor and not vanishing on shell [Ref.  \cite{calmet, Alvarez}]. As already mentioned, a more complete analysis within the so-called low-energy effective theory of gravity approach is presented in \cite{Claudia}. Here, we limit ourselves to consider only the Goroff-Sagnotti term. We begin with the definition of the static metric form defined in equation \eqref{r13}. Then the effective Lagrangian reads
\begin{equation}
 L=\sqrt{\frac{f}{g}}\left(  rg'+g-1+\frac{\omega}{r^4}H^3 \right)\,,   
\end{equation}
where the first term is the usual GR contribution, while the second term comes from the Goroff-Sagnotti invariant, with $\omega $ GS coupling constant and  reads  \cite{Alvarez}
\begin{equation}
 H=r^2\frac{gf'^2}{f^2}+2r\frac{gf'}{f}-2r^2\frac{gf''}{f}-4g+4+2rg'-r^2\frac{f'g'}{f}\,.
\end{equation}

The above Lagrangian is an higher-order differential Lagrangian $L=L(f,g,f',g',f'')$, and the equations of motion are
\begin{subequations}
\begin{eqnarray}
\frac{1}{2\sqrt{fg}}\left(rg'+g-1+\frac{\omega}{r^4}H^3 \right)+ \frac{3\omega\sqrt{f}H^2}{\sqrt{g}r^4} \partial_f H-\frac{d}{dr}\left(\frac{3\omega\sqrt{f}H^2}{\sqrt{g}r^4} \partial_{f'} H \right) + \frac{d^2}{dr^2}\left(\frac{3\omega\sqrt{f}H^2}{\sqrt{g}r^4} \partial_{f''} H\right)=0\,,
\label{q1}
\end{eqnarray}
\begin{eqnarray}
-\frac{\sqrt{f}}{2\sqrt{g}g}\left(  rg'+g-1+\frac{\omega}{r^4}H^3 \right)+ \frac{\sqrt{f}}{\sqrt{g}}\left(1+3\omega\frac{H^2}{r^4} \partial_g H\right)-\frac{d}{dr}\left(r\sqrt{\frac{f}{g}}+  \frac{3\omega\sqrt{f}H^2}{\sqrt{g}r^4} \partial_{g'} H \right)=0\,.
\label{q2}
\end{eqnarray}
\end{subequations}

It is not possible to find the exact solution of these differential equations. One may try to look for a solution when the coupling constant $\omega$ is small. 
Thus
\begin{equation}
f(r)=f_0(r)+\omega f_1(r)+\omega^2f_2(r)+...\,, \quad     g(r)=g_0(r)+\omega g_1(r)+\omega^2g_{2}(r)+...\,.
\end{equation}

The zero order is GR, namely when $\omega=0$, one has 
\begin{subequations}
\begin{equation}
\frac{1}{2\sqrt{f_0g_0}}\left(  rg_0'+g_0-1 \right)=0\,,
\end{equation}
\begin{equation}
-\frac{\sqrt{f_0}}{2\sqrt{g_0}g_0}\left(rg_0'+g_0-1 \right)+\frac{\sqrt{f_0}}{\sqrt{g_0}}-\frac{d}{dr}\left(r\frac{\sqrt{f_0}}{\sqrt{g_0}}\right)=0\,.
\end{equation}     
\end{subequations}

The solutions are 
\begin{equation}\label{f0g0}
 g_0=f_0=1-\frac{2M}{r}\,,   
\end{equation}
with $M$ be the constant of integration. Making use of equation \eqref{f0g0}, one can get the following set of equations
\begin{subequations}
\begin{eqnarray}
    \frac{r^7 \left(r g_{1}'+g_{1}\right)+6912 M^2 \left(2 M-15 (r-2 M)\right)}{2 r^6 (r-2 M)}\omega  +\mathcal{O}\left(\omega ^2\right)=0\,,
\end{eqnarray}
\begin{eqnarray}
    \frac{(2 M-r) \left(r^8 f_{1}'+6912 M^2 (8 M-3 r)\right)+2 M r^7 f_{1}+r^8 (-g_{1})}{2 r^6 (r-2 M)^2}\omega+\mathcal{O}\left(\omega ^2\right)=0\,.
\end{eqnarray}    
\end{subequations}

On solving the above equations, one can obtain the values for $g_{1}(r)$ and $f_{1}(r)$ as follows:
\begin{subequations}
\begin{equation}
 g_1(r)= 2^8 \frac{9 M^2}{r^6}\left(\frac{16 M}{r}-9  \right)\,,
 \end{equation}
\begin{equation}
f_{1}(r)=2^{8}\frac{9M^{2}}{r^{6}}\left(\frac{4M}{r}-3\right)\,.
\end{equation}    
\end{subequations}
 
We can express the equations for $g(r)$ and $f(r)$ as:
\begin{subequations}
 \begin{equation}
 g(r)=1-\frac{2M}{r}+\omega   2^8 \frac{9 M^2}{r^7}\left(-2M+9(2M-r)  \right)\,, 
\end{equation}
\begin{equation}
 f(r)=1-\frac{2M}{r}+\omega   2^8 \frac{9 M^2}{r^7}\left(-2M+3(2M-r)  \right)\,. 
\end{equation}
\end{subequations}

If we take the difference of the above two equations, we get
\begin{equation}
    f(r)-g(r)=-\omega   2^8 \frac{54 M^2}{r^7}\left(2 M-r  \right) \,.
\end{equation}

On the horizon $g(r_H)=0$.  This gives  $r_H=2M+\omega r_1+...$,  with $r_1=\frac{72}{M^3}$. As a result, $f(r_H)=0$, and, to this first-order in $\omega$,  one is dealing with a BH.

Up to the second order, one gets 

\begin{subequations}
\begin{eqnarray}
    \frac{2^{17}3^{6} M^3 \left(2^{2}683 M^2-2319 M r+480 r^2\right)-r^{13} \left(r g_{2}'+g_{2}\right)}{2 r^{12} (2 M-r)}\omega ^2+\mathcal{O}\left(\omega ^3\right)=0, \label{diff_FO_g}
\end{eqnarray}
\begin{eqnarray}
-\frac{2^{17}3^{5} M^{3} \left(2^{4}123 M^3- 2^{4}155M^{2} r+3^{2}115 r^2 M -2^{4}3^{2}r^{3}\right)+[r^{14} \left((r-2 M) f_{2}'+g_{2}\right)-2 M r^{13} f_{2}]}{2 r^{12} (r-2 M)^2}\omega ^2+\mathcal{O}\left(\omega ^3\right)=0\,.~\nonumber\\ \label{diff_FO_f}
\end{eqnarray}
\end{subequations}

Upon solving the aforementioned differential equations \eqref{diff_FO_g} and \eqref{diff_FO_f}, the equations for $g_{2}(r)$ and $f_{2}(r)$ can be expressed as follows:
\begin{subequations}
  \begin{eqnarray}
g_{2}(r)&=&-\frac{2^{17}3^{5} M^{3} \left(7513 M^2-3^{2}773 M r+2^{4}3^{2}11 r^2\right)}{11 r^{13}}\,,
\end{eqnarray}
\begin{eqnarray}
    f_{2}(r)=-\frac{2^{16}3^{5} M^3 \left(2^{2}451 M^2-3^{2}233 M r+576 r^2\right)}{11 r^{13}}\,.
\end{eqnarray}
\end{subequations}

Furthermore,  in order to check if one is dealing with a BH or WH, let us compute up to the second order in $\omega$,  the difference  $f(r_H)-g(r_H)$, with  $r_H=2M+\omega r_1+\omega^2r_2+..$.
\begin{subequations}
\begin{equation}
 f(r)=1-\frac{2M}{r}+ \omega 2^8 \frac{9 M^2}{r^6}\left(\frac{4 M}{r}-3  \right)+\omega^2 f_2(r)\,, 
\end{equation}
\begin{equation}
 g(r)=1-\frac{2M}{r}+ \omega 2^8 \frac{9 M^2}{r^6}\left(\frac{16 M}{r}-9  \right)+\omega^2 g_2(r)\,. 
\end{equation}    
\end{subequations}

Thus, 
\begin{equation}
 f(r)-g(r)=- \omega   2^8 \frac{54 M^2}{r^6}\left(\frac{2 M}{r}-1 \right)+\omega^2 (f_2(r)-g_2(r))\,, 
\end{equation}
and $g(r_H)=0$
\begin{eqnarray*}
    r_H=2M-\omega 2^8 \frac{9M^2}{r_{H}^6}\left(16M-9r_H\right)-\omega^2 r_H g_2(r_H)\,,
\end{eqnarray*}
\begin{equation}
 f(r_H)-g(r_H)= \omega^2  \frac{108}{M^5}r_1+\omega^2 (f_2(r_H)-g_2(r_H))\,. 
\end{equation}

We have
\begin{equation}
 f(r_H)-g(r_H)= \omega^2  \frac{108}{M^5}\left(\frac{72}{M^3}  \right)+\omega^2 (f_2(2M)-g_2(2M))\,. 
\end{equation}

Since
\begin{eqnarray}
f_{2}(2M)=\frac{2^{4}\times 3^{5}\times 43}{11 M^8}\,, ~~~g_2(2M)=\frac{2^{4}\times 3^{5} \times 65}{11 M^8}\,,   
\end{eqnarray}
\begin{eqnarray}
f_{2}(2M)-g_2(2M)=-\frac{2^{5}\times 3^{5}}{M^8}\,.
\end{eqnarray}

One gets
\begin{equation}
 f(r_H)-g(r_H)=0+\mathcal{O}(\omega^3)\,.   
\end{equation}

At the second order in $\omega$,  since $g(r_H)=0$,  it follows that $f(r_H)=0$ and one is dealing with a BH.

Some remarks: First, the second-order corrected metric presented above is the main result of our paper. 
Up to the first order in $\omega$,  our expressions for the metric are in agreement with several others works.  First, Calmet et al \cite{calmet} have found
\begin{equation}
f(r)=1-\frac{2M}{r}+a\frac{M^3}{r^7}\,, \quad g(r)=1-\frac{2M}{r}+b\frac{M^2}{r^6}\left(27-\frac{49M}{r}\right)\,,
\end{equation}
where $a=640 \pi \omega$ and $b=128\pi \omega$, with $\omega$ a small positive constant which describes the quantum corrections, namely for $\omega=0$,  $a,b=0$, and one recovers the classical result. The horizon is determined by  the equation

\begin{equation}
1-\frac{2M}{r_H}+b\frac{M^2}{r_H^6}\left(27-\frac{49M}{r_H}\right)=0\,,
\label{t1}
\end{equation}
which can be solved numerically or in some approximation. In fact, putting $r_H=2M+\omega r_1$ and working with small $\omega$, one has
\begin{equation}
r_H=2M -\frac{10 \pi  }{M^3}\omega+ ..\,. 
\end{equation}

Now, if we compute $f(r_H)$, one has  
\begin{equation}
f(r_H)=-\frac{5\pi \omega  }{M^4}+ \pi \omega\frac{5}{M^4}=0\,.  
\end{equation}

At first order in $\omega$, we have put $r_H=2M$.  Furthermore,  in a recent paper \cite{Sauer}, the Goroff-Sagnotti modification of GR has been reconsidered. At first order in the coupling constant, their result reads  (here $r_s=2M$)
\begin{equation}
  f(r)=1-\frac{r_s}{r}+2\lambda\frac{r_s^2}{r^6}\left(3-2\frac{r_s}{r}\right)\,, \quad g(r)=  1-\frac{r_s}{r}+2\lambda\frac{r_s^2}{r^6}\left(9-8\frac{r_s}{r}\right) \,.
\end{equation}

This result is in agreement with an older one \cite{Anselmi}.   The horizon is 
\begin{equation}
 r_H=r_s-2\lambda\frac{r_s^2}{r_H^5}\left(9-8\frac{r_s}{r_H}\right)=r_s-\frac{2\lambda}{r_s^3}+\mathcal{O}\left(\lambda^2\right)\,.   
\end{equation}

As a result, $f(r_H)=0$.

\section{\texorpdfstring{$F(R)$}{} modified gravitational models}\label{GR_MGM}
It is known the important role of the Starobinsky model in order to describe Inflation \cite{Staro}. Recently, the Starobinsky model has been extended, adding further quadratic terms in curvature tensors, inspired by String Theory, the so-called Starobinsky-Bel-Robinson model \cite{Ketov}. Both the inflation  \cite{Ketov1} and BH solutions have been considered \cite{Ketov2}.

In this section, we investigate a class of modified gravity models which include the original Starobinsky model in a generic SSS. As well known, no exact solutions are at disposal, and the Frobenius method and numerical analysis have been used (see, for example, \cite{bu, Bonanno})  in order to investigate solutions making use of an expansion around a horizon. The  BH solutions that have been found are, in some sense, quite different from the Schwarzschild ones. Also, WH solutions have been investigated. 

Here, as done previously, we shall investigate approximate solutions, which are small deviations from the Scwharzschild solution. We shall commence with the static metric as defined in equation \eqref{r13}. We shall deal again with a general case of modified gravity of the type $F(R)$ where Einstein-Hilbert Lagrangian is present, but in addition, higher-order curvature terms depending only on the Ricci scalar are added. The simplest example is the well-known Starobinsky model for inflation \cite{Staro} $F(R)=\gamma R+\omega R^2$, $\gamma=M^2_P$ Planck mass, and $\omega$ a dimensionless parameter.

As a second example, we have the  Dark Energy Starobinsky model \cite{Staro1}, with
\begin{equation}
 F(R)=\gamma R+R_*^2\left(\left(1+\frac{R^2}{R_*^2}\right)^{-n}-1  \right)  \,.
\end{equation}

A third is the generalization of the Dark Energy model by Hu and Sawicki \cite{Hu}
\begin{equation}
 F(R)=\gamma R-\frac{a R^{2n}}{1+ aR^{2n}} \,.  
\end{equation}

Finally, another example is a generalization of the Dark Energy exponential model \cite{Sebastiani1}
\begin{equation}
 F(R)=\gamma R+a^2\big(1-e^{\frac{-R^2}{a^2}}\big)\,.   
\end{equation}

Other examples can be found in \cite{Giacomozzi}.

\subsection{Equations of Motion for Modified  \texorpdfstring{$F(R)$}{} Gravitational Models}
Here, the following technique presented by the Ref. \cite{Saffari} is valid for SSS. We will investigate mainly the Starobinsky model within  SSSs, but the generalization to generic modified gravitational models is straightforward. We introduce the quantity
\begin{equation}
    X=\frac{f(r)}{g(r)}\,,
\end{equation}
then the Ricci scalar can be expressed as
\begin{equation}
    R=\frac{2}{r^2}-\frac{1}{X}\left(f''+\frac{2f+4rf'}{r^2}-\frac{X'}{X}\left[\frac{f'}{2}+\frac{2f}{r} \right]  \right)\,.
\label{c}
\end{equation}

Thus, as variables, we may take $X$ and $f$. 

\begin{subequations}
\begin{equation}
    \frac{X'}{X}\left(2F_{R}+rF_{R}' \right)-2rF_{R}''=0\,,
\label{e1}
\end{equation}
\begin{equation}
r^2f''-2f+2X +\left(rf'-2f \right) \left( \frac{r F_{R}' }{F_{R}}-\frac{rX'}{2X} \right)=0\,.
\label{e2}
\end{equation}    
\end{subequations}

We have checked that these equations of motion reproduce the exact nontrivial BH solutions discussed in \cite{Sebastiani} within another approach. Recall that   $F(R)=\gamma R+\omega R^2$, $\gamma=M^2_P$ .  When $\omega=0$, one has GR, with ${(F_{R})}_0=\gamma$ constant. The first equation gives $X_0=C_0$ constant. The second equation leads to 
\begin{equation}
    r^2 f_0''-2f_0+2C_0=0\,.
\end{equation}

We know that in GR, the unique solution is $f_0=1-\frac{r_s}{r}=g_0$. As a result, $C_0=X_0=1$. We can now look for approximate solutions for small $\omega$ of equations of motion (\ref{e1})  and (\ref{e2}). An expansion around the GR solution in the power series of the small coupling $\omega$ up to the second-order
\begin{equation}
 f=f_0+\omega f_1+\omega^2 f_2+...\,,\quad X=1+\omega X_1+\omega^2 X_2+...\,.   
\end{equation}

We expand also the Ricci scalar, recalling that $R_0=0$, namely 
\begin{equation}
R=\omega R_1+\omega^{2} R_2+...
\end{equation}

Up to the second order in $\omega$, the equation (\ref{e1})  leads to
\begin{equation}
   X'_1=0\,,\quad \gamma X_2'=2rR_1''\,. 
\end{equation}

Thus,
\begin{equation}
    X_1=C_1\,,\quad X_2=\frac{2}{\gamma}\left(rR'_1-R_1  \right)+C_2 \,.
\end{equation}

On the other hand, one has
\begin{equation}
    X_1=\frac{f_1-g_1}{f_0}\,,\quad X_2=\frac{f_0f_2-f_1g_1-g_2f_0+g_1^2}{f_0^2}\,.
\end{equation}

As a consequence
\begin{equation}
    f_1-g_1=C_1f_0\,,\quad g_2=f_2-f_0X_2-g_1C_1\,.
\end{equation}

Making use of equation (\ref{e2}), one has 
\begin{equation}
    r^2f_1''-2f_1+2C_1=0\,.
\end{equation}

The general solution is
\begin{equation}
    f_1=C_1+b_1r^2+\frac{b_2}{r}\,.
\end{equation}

In order to have an asymptotically Minkoskian flat solution, one may take $b_1=0$ and $C_1=0$. Thus, $X_1=0$, and one also have $g_1=f_1=\frac{b_2}{r}$. At this first order, the solution is still Schwarzschild one, with a small correction in the mass
\begin{equation}
 f(r)=g(r)=1-\frac{r_s-\omega b_2}{r}+..\,,  
\end{equation}
from above equation and equation (\ref{c}), one obtains
\begin{equation}
    R=\frac{2}{r^2}\omega+...\,,\quad R_1=\frac{2}{r^2}\,,
\end{equation}
and 

\begin{equation}
    X_2=-\frac{8}{\gamma r^2}+C_2\,.
\end{equation}

At the second order in $\omega$, equation (\ref{e2})  gives
\begin{equation}
    r^2f_2''-2f_2=-2C_2-\frac{24}{\gamma r^2}+\frac{60 r_s}{\gamma r^3}\,.
\end{equation}

The general solution is
\begin{equation}
  f_2=b_4r^2+\frac{b_5}{r}+C_2-\frac{6}{\gamma r^2}+\frac{6r_s}{\gamma r^3}\,.  
\end{equation}

If we require an asymptotic flat solution, we may take $C_2=b_4=0$. Thus
\begin{subequations}
\begin{equation}
  f(r)=1-\frac{r_s-\omega b_2-\omega^2 b_5}{r}-\omega^2\frac{6 }{\gamma r^2}\left(1-\frac{r_s}{ r}\right)\,,  
\end{equation}
\begin{equation}
    g(r)=1-\frac{r_s -\omega b_2-\omega^2 b_5}{r}+\omega^2 \frac{2}{\gamma r^2}\left(1-\frac{r_s}{r}\right)\,.
\end{equation}    
\end{subequations}

Now, we shall discuss the existence of a horizon.  One can obtained $r_H$ solving $g(r_H)=0$, namely

\begin{equation}
  r_H=r_s-\omega b_2-\omega^2 b_5- \omega^2\frac{2}{\gamma r_H^2}(r_H-r_s  )\,.  
\end{equation}

The approximate solution is
\begin{equation}
    r_H=r_s-\omega b_2-\omega^2b_5+\mathcal{O}(\omega^3)\,.
\end{equation}

In order to understand the nature of the solution, we have to evaluate $f(r_H)$.  One gets
\begin{equation}
    f(r_H)=0+\mathcal{O}(\omega^3)\,.
\end{equation}

Thus, at this order, one again has a BH.

\section{Photon Rings and Shadows}\label{PRS}

In this section, we will investigate the concept of photon ring and shadow associated with a generic SSS compact object, namely a BH or WH.  The important of this issue relies on the fact that strong gravity effects occur there. The topics are standard and can be found in several papers and textbooks (see, for example, \cite{Perlick, Sunny22} and references therein).  For the sake of completeness, we present, in the following, the main results.

To be sufficiently general, we make use of a generic SSS space-time metric, namely
\begin{equation}
ds^2=g_{tt}dt^2+2g_{t\rho}dt d\rho+g_{\rho \rho}d \rho^2+r(\rho)^2\left(d\theta^2+\sin \theta^2 d \phi^2  \right)    \,.
\end{equation}

The well-known parametrization invariant approach allows one to write down as Lagrangian related to a test particle  the quantity
\begin{equation}
 \mathcal{L}=\frac{\dot{s}^2}{2}=\frac{1}{2}\left( g_{tt}\dot{t}^2+2g_{t\rho}\dot{t} \dot{\rho}+g_{\rho \rho}\dot{ \rho}^2+r(\rho)^2\left(\dot{\theta}^2+\sin^2 \theta \dot{ \phi}^2  \right)\right)    \,,
\end{equation}
together, the normalized four-velocity
\begin{equation}
  g_{\mu \nu}\dot{x^\mu}\dot{x^\nu}=k \,, 
\label{ka}
\end{equation}
where $\dot{A}=\frac{d A}{d \lambda}$, with $\lambda$ an affine parameter, and $k=0$ for mass-less particle and $k=-1$ for massive one. In this case, the affine parameter coincides with the proper time. From the above Lagrangian, one gets two conserved quantities
\begin{equation}
   E=g_{tt}\dot{t}+g_{t\rho}\dot{\rho}\,,\quad L_{\phi}=r^2\sin^2 \theta \dot{\phi}\,. 
\label{f1}
\end{equation}

Making use of the equation of motion for $\theta$, namely
\begin{equation}
  \frac{d}{d \lambda}L_\theta=r^2 \sin \theta \cos \theta \dot{\phi}^2\,,  \quad L_\theta=r^2\dot{\theta}\,,
\end{equation}
and  the expressions for the conserved quantities $E$ and $L_\phi$, it follows that the total angular momentum 
   \begin{equation}
  L^2=L_\theta^2+\frac{L_\phi^2}{\sin^2 \theta}\,,     
\label{f2}
\end{equation}
is also conserved. Finally, the first integral related to $\rho$ follows from (\ref{ka}). Thus
\begin{equation}
\dot{\rho}^2=\frac{g_{tt}}{(g_{t\rho}^{2}-g_{tt}g_{\rho \rho})}\left(\frac{E^2}{g_{tt}}+\frac{L^2}{r^2}-k   \right)\,.    
\label{f3}
\end{equation}

Making use of the two first integrals (\ref{f1})  and (\ref{f2}), the Lagrangian becomes

\begin{equation}
 \mathcal{L}=\frac{1}{2}\left( g_{\rho \rho}-\frac{g_{t\rho}^2}{g_{tt}}\right)\dot{ \rho}^2+\frac{1}{2}\left( \frac{E^2}{g_{tt}}+\frac{L^2}{r^2(\rho)}  \right)  \,. 
\end{equation}

As a result, the equation of motion for $\rho$
\begin{equation}
 \frac{d}{d \lambda}\left[ \left( g_{\rho \rho}-\frac{g_{t\rho}^2}{g_{tt}}\right)\dot{\rho}   \right]=-\frac{\mathring{g_{tt}}E^2}{2 g_{tt}^2}-\frac{L^2\mathring{r}}{r^3}  \,. 
\label{f4}
\end{equation}

For circular orbits, $\dot{\rho}=0$. Thus,  from equations (\ref{f3}) and (\ref{f4}), one has 
\begin{equation}
  E^2+\frac{L^2 g_{tt}}{r^2}=k g_{tt}\,,\quad \frac{E^2 \mathring{g_{tt}}}{2 g_{tt}^2}=-\frac{L^2 \mathring{r}}{r^3}\,.  
\end{equation}

We may put $g_{tt}=-f(\rho)$. Thus, in agreement with (see for example \cite{Perlick} and references therein)
one gets
\begin{equation}
 E^2(r_c\mathring{f}_c-2\mathring{r_c}f_c)=2kf_c^2\,, \quad b^2=\frac{L^2}{E^2}=\frac{r_c^3 \mathring{f_c}}{2f_c^2}\,,   
\end{equation}
where we have introduced the impact parameter $b=\frac{L}{E}$. For a mass-less particle ($k=0$),  one gets  
\begin{equation}
 (r_c\mathring{f_c}-2\mathring{r_c}f_c)=0\,, \quad b^2=\frac{r_c^2 }{f_c}\,.    
\label{pr}
\end{equation}

These two equations permit the computation of $r_c$ and $b$ for mass-less particles. For example, for the Scwarschild BH, one has $r=\rho$, $f=1-\frac{r_s}{r}$,  $r_c=\frac{3r_s}{2}$, $b=\frac{3\sqrt{3}}{2}r_s$.

With regard to the shadow, the angle $\alpha$  from the light ray, starting from $r_o$ and the radial direction is given by \cite{Perlick}

\begin{equation}
 (\cot \alpha)^2=\frac{g_{\rho\rho}}{r^2}\left(\frac{d \rho}{d \phi}\right)^2   \,,
\end{equation}
in which the quantities on the right side are evaluated at $r_o$. Making use of the equation of motion for $\rho$, one gets
\begin{equation}
  \sin^2 \alpha=\frac{b^2 f_o}{r^2_o}\,.  
\end{equation}

For $r_o$ very large and for asymptotically flat metrics, one arrives at 
\begin{equation}
 \alpha \simeq \frac{r_c}{r_o\sqrt{f_c} }\,.   
\end{equation}

In the case of Schwarzschild BH, one obtains the well known result
\begin{equation}
    \alpha \simeq \frac{3\sqrt{3}M}{ r_o}\,.
\end{equation}

WHs may be solutions of GR with exotic matter, but as discussed in the previous sections,  solutions of modified gravity models or corrections to GR are also available. WHs admit photon spheres, which then possess shadows. For example, the Damour-Soludhukin WH has been investigated in \cite{Sunny22}. As a first example of WH, let us consider now the $R^2$  static WH solution with $\rho=r$ and 
\begin{equation}
  f(r)=\frac{1}{(1+\lambda)^2}\left(\lambda+\sqrt{1-\frac{r_s}{ r}}  \right)^2\,, \quad g(r)=1-\frac{r_s}{ r}\,.  
\end{equation}

We recall that this WH solution is asymptotically flat,  it reduces to the Schwarzschild BH for $\lambda=0$, and since $f(r)=1-\frac{r_s}{(1+\lambda)r}$ for large $r$, one may interpret $2M=\frac{r_s}{1+\lambda}$  as the mass of this BH mimicker. 

In the following, we compute the photon ring and the shadow of this compact object, its throat being 
$r_H=r_s$.  The photon ring can be evaluated by making use of equation \eqref{pr}. Thus

\begin{equation}
    r_c=\frac{3}{2}r_s-\lambda \sqrt{r_c^2-r_cr_s}\,.
\end{equation}

For $\lambda=0$, one recovers the GR result $r_c=\frac{3}{2}r_s$. Assuming $\lambda$ is very small,  one has
\begin{equation}
r_c=\left( \frac{3-\lambda\sqrt{3}}{2}\right) r_s\,.
\end{equation}
 
With regard to the angle shadow, one has

\begin{equation}
    \alpha \simeq \frac{3\sqrt{3}M}{ r_o}\left(1+\left(2-\frac{2}{\sqrt{3}}\right)\lambda  \right)\,.
\end{equation}

This result is valid for very small $\lambda$ for large $r_o$.

Finally, for the black holes solution found  within the effective theory formulation of gravity, we remind that at first order in the generic coupling constant, one has
\begin{equation}
f(r)=1-\frac{2M}{r}+\omega f_1(r)\,, \quad g(r)=1-\frac{2M}{r}+\omega g_1(r)\,,\,\quad g_1(2M)=f_1(2M)\,.
\end{equation}

The photon ring radius is given by
\begin{equation}
 r_c=3M-\omega r_c\left(\frac{r_c f'_1(r_c)}{2}-f_1(r_c) \right)\,.   
\end{equation}

For small $\omega$ one has

\begin{equation}
 r_c \simeq3M\left(1-\omega \left(\frac{3M f'_1(3M)}{2}-f_1(3M)\right)\right)\,.   
\end{equation}

The recent papers have been published, providing bounds on the model parameters $M$. For further details, please refer to the papers by \cite{Collaboration2022, Vagnozzi2023}.

\section{Conclusion}\label{Conclusion}

In this paper, first  we have revisited a non-singular static metric, including black holes not admitting a de Sitter core in the centre and traversable wormhole,  within a class of higher-order $F(R)$ modified gravity, satisfying the constraints $F(0)=\frac{dF}{dR}(0)=0$. These two conditions allow to satisfy the equation of motions, even though a large class of $F(R)$ models are possible. Among them, the pure quadratic gravity model, namely $F(R)=R^2$ is of particular interest.   

In the second part of the paper, making use of the so called effective field theory formulation of gravity, the quantum corrections to GR has been considered.  We recall that within this approach one starts with the Einstein-Hilbert action plus  additive higher-derivative  terms, typically built with curvature invariants, with related arbitrary coupling constants. In our approach, such coupling constants have been considered small enough in order to justify linear and quadratic perturbation  in the coupling constants around the GR solutions.  In the important case of SSSs, it is well known that the zero order solution is the Schwarzschild BH.  In this paper, we have not investigated perturbation around the flat Minkoswki solution. 

 In particular, in the case of  Einstein-Hilbert action  plus cubic curvature Goroff-Sagnotti contribution,   the second order corrections in the Goroff-Sagnotti coupling constant  have been  computed. 
In general, it has been  shown that the effective metrics, namely Schwarschild expression plus small quantum correction are related to  black holes, and  not to traversable wormholes. In this framework, within the approximation considered, the resolution of singularity for $r=0$  has not been resolved.  

Furthermore, since strong gravity effects may occur around the photon rings, a short review of this issue with some applications has been added.   

Finally, with regard to future work, we think that one could try to go beyond the second order approximation in the coupling constant, in order to check if the black hole solutions are still present, or wormholes  solutions arise. The presence of wormholes solution may occur, since the higher-derivative terms can violate WEC (weak energy conditions), as shown in Section \ref{NSBH} and appendix \ref{BWS}. In fact, our formalism is suitable for this task, since, in the SSS case, we start with metric having the function $f(r)$ and $g(r)$ not proportional, namely with the choice $f(r) \neq g(r)b(r) $.

 As far as  the possible presence of WHs  is concerned,  we note that this fact has been numerically verified within Einstein-Weyl gravity (see for example \cite{Bonanno2} and references therein) .


\begin{appendix}

\section{\texorpdfstring{$\mathcal{C}^{n}$}{} theory}
Here, following \cite{Cardoso}, we verify that at the first order of the curvature coupling constant $\varepsilon$, the higher derivative model described by the action
\begin{equation}
  I_n=\int d^4x \sqrt{-g}\left(R-\varepsilon a_n Q^n\ \right)  \,, \quad Q=\sqrt{R_{\mu \nu\alpha \beta}R^{\mu \nu \alpha \beta  }}\,,
\end{equation}
admits BH solution. The case $n=2$ has been investigated in Section \ref{n2}. At first order, one has
\begin{equation}
 f(r)=1-\frac{2M}{r}+\varepsilon \left(\frac{2M}{r}\right)^{6n-2}F_{n}(r) \,,       
\end{equation}
where
\begin{equation}
F_n(r)=-\frac{48^n(n-1)}{6(2n-1)}\left( 1+2n+(12n^2-6n)\left(\frac{r}{M}-2\right) \right)\,, 
\end{equation}
\begin{equation}
g(r)=1-\frac{2M}{r}+\varepsilon \left(\frac{2M}{r}\right)^{6n-2}G_n(r) \,,  
\end{equation}
\begin{equation}
G_n(r)=-\frac{48^n(n-1)}{6(2n-1)}\left(1-6n+4n \frac{r}{M} \right)\,.
\end{equation}

At first order, the horizon is
\begin{equation}
 r_H=2M\left(1+\varepsilon G_n(2M) \right)\,.   
\end{equation}

Evaluation of $f(r)$  on the horizon gives
\begin{equation}
  f(r_H)=\varepsilon \left(G_n(2M)-F_n(2M) \right)=0\,.  
\end{equation}

As a result, at this order, one has a BH solution.

\section{A specific class of modified gravitational models}

In this appendix,  we make use of another method to investigate the modified gravitational of Section \ref{GR_MGM}.
 Then the effective Lagrangian reads
\begin{equation}
 L=\sqrt{\frac{f}{g}}\left(  rg'+g-1+\omega H(r^2,f,g,f',g',f'',g'') \right)\,,   
\end{equation}
where the first term is the usual GR contribution, while the second term comes from higher curvature invariants, with $\omega $ GS coupling constant. In the case of Starobinsky model  $H=r^2 R^2$, with $R$ the Ricci scalar and $\omega=\frac{1}{6M^2}$, $M$ a suitable but large mass scale.

 As we already stressed, the crucial point is that when $\omega=0$, one has to deal with GR, and since we are in the vacuum, the only solution is the Schwarzschild BH, namely
\begin{equation}
f_0(r)=g_0(r)=1-\frac{r_s}{r}\, .    
\end{equation}

Thus, one may make the expansion 
\begin{equation}
f(r)=f_0+\omega f_1+\omega^2 f_2+..\,, ~~~g(r)=g_0+\omega g_1+\omega^2 g_2+..\,, ~~~H=H_0+\omega  H_1+\omega^2 H_2+...\,.
\end{equation}

As a result, one has at the second order in $\omega$

 \begin{eqnarray}
  L&=& r g_1' +g_1+H_0+ \omega \frac{f_1-g_1}{2g_0}\left( rg_1'+g_1+H_0 \right)+\omega(rg_2'+g_2+ H_1)+ \omega^2\left( \frac{f_1-g_1}{2g_0}\left( rg_2'+g_2+H_1 \right)\right) \nonumber \\
&&+\omega^2\left(\frac{4f_2g_0-4g_2g_0-2f_1g_1-f_1^2+3g_1^2}{8g_0^2}\right)\left( rg_1'+g_1+H_0 \right)+\omega^2 \left( rg_3'+g_3+H_2 \right)+\mathcal{O} (\omega^3)\,.
\end{eqnarray} 

First, one may note that there exists a simplification in the cases $H_0=0$ and $H_1=0$. In fact, in this case, one gets

\begin{eqnarray}
    L&=& r g_1' +g_1+\omega \frac{f_1-g_1}{2g_0}\left( rg_1'+g_1 \right)+\omega(rg_2'+g_2)+ \omega^2 \frac{f_1-g_1}{2g_0}\left( rg_2'+g_2 \right)    \nonumber \\
&&+\omega^2 \left(\frac{4f_2g_0-4g_2 g_0-2g_1f_1-f_{1}^{2}+3g_1^2}{8g_0^2} \right)\left( rg_1'+g_1 \right)+\omega^2(rg_3'+g_3+H_2)+\mathcal{O}(\omega^3)\,.
\end{eqnarray}
 
For example, in the case of the Starobinsky inflationary model, $H=r^2R^2$. Thus 
\begin{equation}
 H=r^2(R_0^2+2\omega R_0 R_1+\omega^2 R_2)\,,\quad H_0=r^2 R^2_0\,, \quad H_1=2r^2 R_0 R_1\,, \quad H_2=r^2 R_1^2 \,. 
\end{equation}

As a consequence, since $R_0=0$, one has $H_0=0$ and $H_1=0$. 

However, there exist in literature several models with this property. They all belong to the class of modified models such that
\begin{equation}
H=r^2f(R)\,, \quad f(0)=0\,,   \quad  f'(0)=0\,.
\end{equation}

For small $\omega$, one has
\begin{equation}
 H=r^2f(R_0+\omega R_1)=r^2f(R_0)+r^2\omega R_1 f'(R_0)+r^2\omega^2\frac{f''(R_0)}{2}R_1^2+...\,.   
\end{equation}

Thus $H_0=r^2f(R_0)\,,~ H_1=r^2f'(R_0)R_1\,,~ H_2=r^2\frac{f''(R_0)}{2}R_1^2 $. In the Scwharzschild case, $R_0=0$, and $H_0=H_1=0$. 

Let us investigate the associated equations of motion. At  the first order corrections in $\omega$,  making the variation with respect to $f_1$, one gets
\begin{equation}
 rg'_1+g_1=0\,.   
\end{equation}

The solution is
\begin{equation}
 g_1=\frac{c_1}{r}\,.   
\end{equation}

The variation with respect to $g_1$, again  at the first order in $\omega$ , leads to
\begin{equation}
 \frac{d}{dr}\left( \frac{r(f_1-g_1)}{g_0}\right)=\frac{f_1-g_1}{g_0}\,.   
\end{equation}

Thus
\begin{equation}
 f_1=g_1+c_2 g_0   \,.
\end{equation}

Thus, at first order in $\omega$,  one has
\begin{equation}
 f(r)=1-\frac{r_s-\omega c_1+\omega c_2 r_s}{r}+\omega c_2\,,\quad    g(r)=1-\frac{r_s-\omega c_1}{r}\,.
\end{equation}
 
 First, we may redefine the time in order to deal with an asymptotic flat Minkowski space-time, $dT^2=(1+\omega c_2 )dt^2\, c_2>0,~ \omega>0  $. Thus we arrive at 
\begin{equation}
 ds^2=-F(r)dT^2+\frac{dr^2}{g(r)}+r^2dS^2  \,, 
\end{equation}
with 
\begin{equation}
 F(r)=1-\frac{1}{r}\left(r_s-\frac{\omega c_1}{1+\omega c_2}\right)\,, \quad g(r)=1-\frac{1}{r}(r_s-\omega c_1) \,.  
\end{equation}

One has an horizon, $g(r_H)=0$, namely $r_H=r_s-\omega c_1$. Since the new Lapse function $F(r)$ on the horizon reads
\begin{equation}
 F(r_H)=-\frac{\omega^2c_1c_2}{(1+\omega c_2)r_H}\,.
\end{equation}

Which has to be interpreted as vanishing at first order in $\omega$. Thus one concludes that we do not have a generalized Damour-Solodhukin WH \cite{Damour} as soon as $c_1<0$, but one is dealing with a Schwarschild-like BH. In fact in the linear $\omega$ approximation, we have 
\begin{equation}
 F(r)=1-\frac{1}{r}\left(r_s-\omega c_1\right)\,, \quad g(r)=1-\frac{1}{r}(r_s-\omega c_1) \,.  
\end{equation}

Thus, we are left with $f_1=g_1=\frac{c_1}{r}$. At the second order in $\omega$,  one has
\begin{equation}
 r g_2'+g_2=4g_0\left(-r^2 R_1 \frac{\partial{R_1}}{\partial f_1}+\frac{d}{dr}\left(r^2 R_1 \frac{\partial{R_1}}{\partial f_1'}\right)- \frac{d^2}{dr^2}\left(r^2 R_1 \frac{\partial{R_1}}{\partial f_1''}\right)  \right)=G_2(r)\,.   
\end{equation}

Thus
\begin{equation}
 g_2(r)=\frac{c_2}{r}+\frac{1}{r}\int G_2(r)dr\,.   
\end{equation}

The variation with respect to $g_1$ at the second  order in $\omega$ gives

\begin{equation}
 \frac{d}{dr}\left( \frac{r(f_2-g_2)}{g_0}\right)-\frac{f_2-g_2}{g_0}=-G_2(r)+r^2 \frac{\partial R_1}{\partial g_1}-
 \frac{d}{dr }\left(r^2 \frac{\partial R_1}{\partial g_1'}  \right)\,.   
\end{equation}

We note that $f_2-g_2=2K_2g_0$. Where $K_2$ is a solution of the differential equation. Thus,  at the second order in $\omega$,  $f_2=g_2$, and one has to deal with a BH, in agreement with the result of Section \ref{GR_MGM}.

\section{Buchdahl wormhole solution}\label{BWS}
Now we consider the non-trivial SSS solution of pure $R^2$ gravity found in \cite{Nguyen}. In isotropic coordinates, one has
\begin{equation}
 F(\rho)=\left( \frac{4\rho-ar_s}{4\rho+ar_s}\right)^\alpha\,,\quad G(\rho)=\left(\frac{1}{4\rho}\right)^4(4\rho+ar_s)^{2-\beta}(4\rho-ar_s)^{2+\beta}\,,
\label{a5}
\end{equation}
where we have put
\begin{equation}
 a=\sqrt{1+3b^2}\,, \quad \alpha=\frac{2}{a}(b+1)\,, \quad \beta=   \frac{2}{a}(b-1)\,.
\end{equation}

Above,  $b$ is a dimensionless free parameter (the Buchdahl parameter), and $r_s$ is the constant radius. When the Buchdahl parameter is vanishing, $a=1$, $\alpha=2$, $\beta=-2$ and one recovers the Scwarzschild solution (\ref{sc}).
The areal radius reads
\begin{equation}
 r(\rho)=\left(\frac{1}{16\rho}\right)(4\rho+ar_s)^{1-\beta/2}(4\rho-a r_s)^{1+\beta/2}\, .
\end{equation}

The location of the horizon is given by solving the equation (\ref{a2}), or alternatively $(\partial_\rho r)^2=0$. Thus, one has
\begin{equation}
 1=\rho_H\left(\frac{4+2\beta}{4\rho_H-ar_s}+ \frac{4-2\beta}{4\rho_H+ar_s}  \right). 
\end{equation}

This leads to
\begin{equation}
16\rho_H^2+4a\beta r_s \rho_H+a^{2}r_s^2=0\,.
\end{equation}

In terms of Buchdahl parameter, one has
\begin{equation}
\rho_H=\frac{r_s}{4}\left(1-b \pm \sqrt{-2b(b+1)}  \right)   \,.
\end{equation}

One can see the role of the sign of $b$. For example, for $b<0$, and $|b|<1$, one may have 
\begin{equation}
\rho_H=\frac{r_s}{4}\left(1+|b| + \sqrt{2|b|(1-|b|)}  \right)   \,.
\end{equation}

Then $\rho_H>0$. In this case, since $F(\rho_H)$ is not vanishing, one is dealing with a WH.

\end{appendix}

\section*{Acknowledgement} BM thanks IUCAA, Pune (India) for providing support for an academic visit during which a part of this work is accomplished. We greatly appreciate the valuable insights provided by the esteem referee.

\end{document}